\documentclass[aps,prd,twocolumn,notitlepage,nofootinbib]{revtex4}
\usepackage{amssymb}
\usepackage{amsmath}
\usepackage{verbatim}
\usepackage{appendix}
\usepackage[pdftex]{graphicx}
\usepackage{color}

\renewcommand{\vr}{\vec{r}}
\newcommand{\vv}{\vec{v}}

\begin{document}

\title{Light deflection in Observation: Angle differences between two null geodesics on the de Sitter spacetime with multi-lensing objects}
\author{Rio Saitou}
\email{riosaitou18@hotmail.com}
\author{Hiromi Saida}
\email{saida@daido-it.ac.jp}
\affiliation{Daido University, Nagoya, Aichi, 457-8530, Japan}

\begin{abstract}
We derive angle differences between two null geodesics, propagating from light sources to an observer, on the de Sitter spacetime with multi-lensing objects.
Assuming the lensing objects are mass monopoles on the de Sitter background, we derive the metric tensor by solving the Einstein equation perturbatively. On that spacetime, we solve a null geodesic parametrized by the coordinate time.
Using the null geodesics, we define the angle differences in a coordinate invariant way.
We take in the relativistic effects up to the first order of perturbation and clarify the magnitude of approximation errors.
%As a consistency check, we confirm the coordinate invariance of the angle differences calculating in another coordinate.
We find that the rest observer, who sees the isotropic cosmic space,
implicitly observes the effect of the cosmological constant on the angle differences through the positions of the light sources.
%up to the first order of perturbation.
As a practical application, we regard the massive black hole at our galactic center (Sgr A*) and the solar system as the lensing objects, further a star and a flare around Sgr A* as the light sources.
%As a practical example, we consider the Sagittarius A*(Sgr A*) and the solar system as the lensing objects and a star and a flare around Sgr A* as the light sources.
We write the angle differences between these light sources using their spatial coordinates.
We find that deflections by Sgr A* remain in the angle differences while deflections by the solar system cancel out up to the first order of perturbation.
The deflections by Sgr A* amounts around 10 microarcseconds, which is detectable in the near future observations.

%We find that up to the first order of perturbation, the cosmological constant does not affect the angle differences for the rest observer at the spatial origin who see the cosmological horizon isotropically.
%
% 2本のヌルがオブザーバーのとこで作る角度差を、レンズ天体＋宇宙定数の時空上で求めた。
% メトリックは、dS上に天体のマスによるポテンシャル置いて摂動的に求めた。
% ヌルを与えられた時空上で、摂動的に解き、座標時によるパラメトリック表示を求めた。
% 相対論効果の１次までを摂動的に取り入れ、誤差も明確にしながら求めた。
% 角度差はゲージ不変な形で定義した。だから観測データと直接比べられる観測量だ。
% 別の座標系でも計算して少なくとも摂動１時まででゲージ不変になってることも確認した。
% 宇宙が東方的に見えるオブザーバーに対しては、摂動１時までだと角度差には宇宙項の効果がないことを見つけた。
% で、実例としてSgr A*周りのS0-2とフレアからくる光の間の角度差を計算して光源やレンズ天体の座標値で具体的に表し、
% １０マイクロくらい補正はいることを見つけた。これは次世代観測で引っかかるレベル。
\end{abstract}

\maketitle

\section{Introduction}

%・ひかりってまがるよねーって話
%我々が観測している光は、蛍光灯の光から遠く離れた銀河から届く電磁波にいたるまで、すべて重力場の影響を受けており、
%軌跡は単純な座標値の差で書かれるような線分にはならない。

Gravitational deflection of light is an important phenomenon predicted by the general relativity and/or other metric theories of gravity.
%When we can ignore the self gravitation of matter and any other forces besides the gravity
When we treat a light or a massless particle just as a point and consider the gravitational force only, we express its path by a geometric straight line, that is, a null geodesic.
To obtain a theoretical prediction for a path of light we observe, we need to solve the null geodesic analytically or numerically for a given geometry.
%
%・これまで計算されてる光の曲がりの話
The gravitational deflection of light is actually detected in a broad physical scale from the deflection by the sun firstly observed by Eddington \cite{Dyson:1920cwa} to the gravitational lensing by the cluster of galaxies in the cosmological scale \cite{lensing}. Among them, it is actively searched that how the cosmological constant affects the light deflection. %Especially, we can solve the null geodesic exactly on the Schwarzschild-de Sitter (SdS) spacetime. でも固有時でパラ目トライズされてるのは解けてないかも。だとすると観測データとは比べにくい。
Especially, with the use of the Gauss-Bonnet's theorem, the deflection angle of light has been exactly defined \cite{Gibbons:2008ru} and derived for the Schwarzschild-de Sitter(SdS) spacetime
and so on \cite{Ishihara:2016vdc,Ono:2019hkw,Takizawa:2020egm}.
%, which is the exact vacuum solution of the Einstein equation with the positive curvature.
%エディントンのから
%重力レンズまでいろんな物理スケールに渡ってたくさん観測されてる。
%そのなかで、宇宙項の影響を調べた研究も活発に行われている。
%特に、Eeqの厳密解であるSdS時空上でヌルもまた厳密に解くことができ、
%厳密に曲がりを定式化している研究も多数あり（浅田さんのとか）。

%・やられてないこととか時空のセットアップとか
In practical observations of the lights, we measure angle differences between the light rays entering the observer.
%So far, coordinate invariant angle differences between two null geodesics are derived for various spacetimes\cite{}.
%
%However, the angle differences on the de Sitter spacetime with multi lensing objects are not analyzed in detail. こういう書き方は怖い。
%しかしながら、de Sitter＋マルチレンズのstatic sliceで、ヌルの角度差がどうなるか、geodesicsを解いて解析的に調べられていない。
Then, how do we derive the angle differences from the null geodesics traveling the present universe?
The present universe is nearer to the de Sitter spacetime with multi lensing objects rather than the SdS spacetime.
%The observer outside the black hole in the SdS spacetime does never observe the cosmological horizon isotropically, so that the SdS spacetime is dissociative from the real universe.
%since the SdS spacetime is constructed putting the black hole in the center of the spacetime.
%In the realistic universe, we accept an observer who sees the cosmological horizon isotropically as a rest observer in the cosmic microwave background (CMB).
Thus, we should consider the spacetime where the lensing objects are put onto the de Sitter background
as the first-order approximation of the real universe.
This construction is similar to the Einstein-Infeld-Hoffman approximation where
the astrophysical objects are put onto the Minkowski background perturbatively.
Instead, we treat the lensing objects as the perturbation on the de Sitter background, as in the cosmological perturbation theory. We note that the de Sitter background referred to here expresses just a local part of the whole universe since by the lights from the astrophysical objects, we can measure the geometrical effect between the observer and those objects only. The local part of spacetime does not involve the whole of our Hubble patch filled by the cosmic microwave background (CMB).
%
%では、角度差をヌルを使って現実の宇宙の時空で計算しようとしたらどうするのだろう？
%・現実の宇宙の時空はSdSではなく、λ＋他の天体も入ったマルチレンズ時空
%・SdS時空におけるBH外部のオブザーバーは宇宙的ホライズンを東方的に見ることはない。これは現実の宇宙膨張と乖離している。それはブラックホールを中心にして考えているからだ。現実の宇宙膨張を考えるならば、CMBに対して静止している人のようにホライズンが東方的に見える観測者を中心に考えるべきで、摂動領域ならばその時空の上に天体がのっかっているという描像のほうが的を射ているだろう。なので、我々は、SdS時空ではなく、de Sitter時空の上に摂動的に天体の影響が乗るような時空を考える。これはEIHでミンコフスキーの上に天体の影響が摂動的に乗っているようなのと同じであり、いわば宇宙論的摂動のように天体の影響を取り扱うという摂動展開だ。
%これが現実の宇宙を表すのに適したセットアップのはずだ。

Also, the actual observational data are obtained at the observer's proper time.
%我々の動きをモデル化した上でなら、固有時を座標時で表すのは難しいことではない。
If we model the motion of the observer properly, it is not difficult to express the proper time by the coordinate time. So, it is useful to solve the null geodesic parametrized by the coordinate time for comparing to the observational data.
This parametrization has a numerical advantage also since we do not need to solve the full geodesic equations but the spatial parts of it only.
%Thus, despite a perturbative approximation, it is very helpful to solve the null geodesic parametrized by the coordinate time for comparing with the observational data.

%・また、実際の角度データはオブザーバーの時間でパラメトライズされてるから、ヌルの理論式も時間でパラメトライズされた形でほしい。
%データはオブザーバーの固有時で測られているが、固有時と座標時の間の変換が分かれば、座標時でパラメトライズされたヌルの理論式はとても使いやすく、有用になる。
%また、数値的にやるときには固有時で全部解くより計算コストが下げられるので、誤差のオーダーが明確にできている限りは有用。
%よって、現実的な宇宙を表す時空において、近似的であれ、解析的に評価できるような時間でパラ目トライズされた便利なヌルの表式が必要。そうして導出されたヌルを使って、ヌルの間の角度差を求めれば、解析解と実際の宇宙における観測データと直接比べられるので、非常に有用となる。
%
%一つ注意しておくと、天体から来る光の観測から測れる宇宙項の影響は、天体とオブザーバーとの間の時空の影響のみであり、そういった意味で、ここで取り入れる宇宙項はローカルな時空についてのものだ。
%これはCMBなどのようにハッブルパッチ全体に関与するものではない。

%さらに、ローカルハッブルの値に注目すると、CMBのように宇宙全体を意識した膨張とは違うローカルな宇宙膨張に注目することができる。
%その場合、宇宙原理などないので、別にCMBとの相対速度とか考えんで良くて、我々を中心としたローカルな宇宙膨張を記述するような時空を考えることもできる。
%特に、銀河系内のオブジェクトを使った観測では、ローカルハッブルしか測ることはできないので、宇宙原理を仮定せずに、そういうローカルなdS時空を考えるほうが適当である。実際、CMBの時代では宇宙は一様等方だったが、構造ができてからの時代では、ローカルハッブルの値が一様等方でないという観測も近年出てきている。なので、系内などの観測に対しては宇宙原理を用いることができるとは必ずしも限らないのだ。（そもそも宇宙原理は作業仮定に過ぎない。）
%この論文では、簡単のため、原点に我々が静止しているようなstatic座標系で、我々を中心とするような(ローカルな時空についての)ホライズンがあると思って、その角度差への影響を調べてみる。このオブザーバーは、flatで見たときには、宇宙が一様東方に見えるようなオブザーバーと一致している。
%つまり、ローカルな当方的な宇宙膨張に対して我々は静止しているオブザーバーであるという状況を考える
%（もし宇宙原理を仮定するならCMBに対して静止しているオブザーバーと考えることもできるが、その場合、我々はCMBに対して静止していないので、我々が観測する観測量には適用できない）。我々がたとえ原点から動いていたとしても、まあそんな大した速度で動いていないなら、短い時間であればそんな大した影響は入らんでしょう。

%・考える物理系

In this article, we derive the angle differences between two light rays which travel in the de Sitter spacetime with multi lensing objects and enter an observer.
We solve the spatial part of the null geodesic parametrized by the coordinate time and use it for the angle differences.
We assume that all the lensing objects move slowly relative to the light speed and treat them as mass monopoles.
This corresponds to the first order post-Newtonian approximation, and we treat the cosmological constant term in the metric as the perturbation also. For all the calculations, we treat the relativistic effect perturbatively and take into the leading order corrections with clarifying the magnitude of the approximation errors.
For simplicity, we regard ourselves as a rest observer at the spatial origin in the static coordinate system. This observer sees the isotropic cosmic space and corresponds to the observer who sees the homogeneous and isotropic universe in the flat slicing of the de Sitter spacetime. We evaluate the angle differences between the light rays measured by this observer for a practical example.

%我々は、複数のレンズ天体と宇宙定数で作られる時空上を測地線にしたがって運動する2本の光を考え、それらの光が観測者の位置で作る角度差を求める。
%レンズ天体はゆっくり動く状況のみを考え、レンズ天体はものポールとして扱える状況に限定する。
%すべての計算は相対論的効果を摂動的に扱い、誤差を明確にしながら、leadingの効果のみ取り入れる。
%一つの重要な結論として、少なくとも摂動の1次のオーダーでは、宇宙が当方的に見えるオブザーバーの位置での角度差に宇宙項の影響は無いということを見る。

%・論文の構成
This article is organized as follows. In Sec.\ref{2}, we solve the null geodesic which travels from a light source to an observer. We expand the solution around a uniform linear motion and take the first-order corrections perturbatively.
In Sec.\ref{3}, we define the angle differences between two null geodesics in a coordinate independent way using the tetrad basis at the observer. We derive an approximate expression for small angle differences by using the apparent velocity of the lights derived in Sec.\ref{2}.
In Sec.\ref{4}, we apply our formula of the small angle differences to a practical example, in which we consider the solar system and the massive black hole at our galactic center (Sagittarius A*, Sgr A*) as the lensing objects.
We express the deflections of lights by the spatial positions of the light sources and the lensing objects. We will find that the deflections by the solar system cancel out for the small angle differences and only the deflection by Sagittarius A* remains.
Sec.\ref{5} is devoted to a summary.
We add three appendices. In App.\ref{A}, we perform a coordinate transformation perturbatively from the usual static slice of the de Sitter spacetime to an isotropic slice.
In App.\ref{B}, we give a technical detail for the derivation of the angle differences.
In App.\ref{C}, we change the coordinate as boosting the rest observer to a moving observer
and we confirm the coordinate invariance of the angle differences calculating in the two different coordinates.
Throughout this article, we normalize the light speed to unity, $c=1$. The spatial components of physical quantities are denoted with arrows or the Roman indices. The dot between the quantities with arrows implies a contraction of their components.
%
%２：測地線の座標時によるパラメータ表示の摂動解
%３：角度差のゲージ不変な定義と、角度差が小さいときの一般的な近似式
%４：小さい角度差の計算の実例。レンズとして銀河中心ブラックホールと太陽系、光源としてＢＨ周りの恒星Ｓ２とフレアを考えて、
%3で導いた式使って実際に角度差への相対論的補正を計算。10マイクロくらいあるから次世代観測で見えそうということを見る。
%５：devoted to summary
%App Aは通常のstatic sliceからisotropicへの変換
%app Bは静止オブザーバーが動いているように見える座標への変換と、本文で定義してる角度差が二つの座標で計算して一致してるかの確認をした。
%
%c=1、矢印つきのはspatialな成分をまとめた量、
%矢印付きの間の演算ドットは単純に成分のかけ算の和。

%%==============================================================
\section{Null geodesics}\label{2}

We derive an explicit solution of null geodesic, propagating from a light source to a timelike observer, in the spacetime with multi-lensing objects and the cosmological constant. We consider a weak gravitational region only, so that we expand the spacetime perturbatively from the Minkowski spacetime. In addition, we assume that all the lensing objects are moving much slower than
light and we ignore all of the multipole moments of the lensing objects.

In this set up, the energy-momentum tensor for slowly moving lensing objects is given as
\begin{align}
T^0_0&= -\rho,\ T^i_0\approx \rho v_J^i,\ T^i_j \approx \rho v_J^iv_{Jj}, \nonumber \\
\rho&\approx \sum_J M_J \delta^3(\vec{r}-\vec{r}_J) \ ,
\end{align}
where $v_J^i$ and $M_J$ are the velocity and the mass of the lensing objects, respectively.
$\vr$ represents the spatial coordinate and $\vr_J$ is the spatial coordinate of the $J$-th lensing objects.
We consider the mass monopoles only for the energy density.
In general relativity, the energy density of mass monopoles generates the gravitational potential treated as the first-order perturbation in the weak field region. The velocity $v_J^i$ of
lensing objects is a small quantity also, and hence we cut off $T^i_0$ and $T^i_j$ as higher-order perturbations
\begin{equation}
\label{ }
T^i_0\approx 0,\ T^i_j \approx 0\ .
\end{equation}
This approximation corresponds to the first order post-Newtonian (1PN) expansion.

Up to the first order of perturbative expansion, the metric has the following form:
%\begin{align}
% ds^2 &= -(1+2\Phi(t,\vr_P) + c_1\Lambda r_{Po}^2)dt^2 \nonumber \\
% &\quad + (1+ 2\Psi(t,\vr_P) + c_2\Lambda r_{Po}^2)d\vec{r}_P^2 \ , \\
% c_1,&\, c_2: {\rm constants} \ ,\nonumber
%\end{align}
%%
\begin{align}
ds^2 &= -(1+2\Phi(t,\vr) )dt^2
+ (1+ 2\Psi(t,\vr) )d\vec{r}^2 \ , \nonumber \\
d\vr^2 &:= dx^2+dy^2+dz^2 ,
\end{align}
%
%where we treat the cosmological constant terms perturbatively as well as the Newtonian potential $\Phi$ and the curvature perturbation $\Psi$.
where we treat the Newtonian potential $\Phi$ and the curvature perturbation $\Psi$ perturbatively.
Note that we have already fixed the gauge, so-called the Newtonian gauge.
The vector and tensor perturbations are ignored since they are nothing more than the higher-order perturbations if the lensing objects and the observer move slowly.
%We obtain the solution of metric from the Einstein equations: $G^\mu_\nu - \Lambda\delta^\mu_\nu = 8\pi G T^\mu_\nu$
We expand the Einstein equations with the cosmological constant up to the first order of perturbations
%\begin{align}
% &2\partial_i\partial_i \Psi + (6c_1+1)\Lambda = -8\pi G \rho, \nonumber \\
% &2\partial_i \dot{\Psi} \approx 0, \nonumber \\
% &-2\ddot{\Psi} +(\delta_{ij}\partial_k\partial_k - \partial_i\partial_j)(\Phi + \Psi)
% \nonumber \\
% &+ [2(c_1+c_2)+1]\Lambda \delta_{ij} \approx 0 \ ,
%\end{align}
%
\begin{align}
\label{Eeq}
&2\partial_i\partial_i \Psi + \Lambda = -8\pi G \rho, \nonumber \\
&\partial_i \dot{\Psi} \approx 0, \nonumber \\
&-2\ddot{\Psi} +(\delta_{ij}\partial_k\partial_k - \partial_i\partial_j)(\Phi + \Psi)
+ \Lambda \delta_{ij} \approx 0 \ ,
\end{align}
where the dot and $\partial_i$ denote the derivatives with respect to the coordinate time $t$
and the spatial coordinates $(x,\,y,\,z)$ respectively.
Since the slowly moving approximation implies
\begin{align}
&\partial_t \sim \frac{v_J}{r}, \nonumber \\
&\dot{\Psi},\ \ddot{\Psi}:{\rm higher\ order}, \nonumber
\end{align}
%
%we obtain a solution
%\begin{align}
% &c_1 = -\frac{1}{6}, \ c_2 = -\frac{1}{3}, \nonumber \\
% &\Phi = - \sum_J \frac{GM_J}{r_{PJ}},\ r_{PJ}:= |\vr_P - \vr_J|, \nonumber \\
% &\Psi = -\Phi \ ,
%\end{align}
%%
%where we chose the Newtonian potential as it falls off at the boundary.
we obtain a solution %with a suitable boundary condition
\begin{align}
&\Phi = - \sum_J \frac{GM_J}{|\vr-\vr_J|} -\frac{1}{6}\Lambda r^2, \nonumber \\
&\Psi = \sum_J \frac{GM_J}{|\vr-\vr_J|} -\frac{1}{12}\Lambda r^2. %\nonumber \\
%&r_{PJ}:= |\vr_P -\vr_J|.%,\quad r_{Po}:= |\vr_P -\vr_o|\ ,
\end{align}
%
%%where $\vr_o$ is the spatial point of the observer.
%%$\vr_o(t)$ is the spatial coordinate of the slowly moving observer and we approximate it by a uniform linear motion:
%%\begin{equation}
%% \label{ro}
%% \vr_o(t) \fallingdotseq \vr_{o,{\rm ini}} + \vv_o t, \quad |\vv_o|\ll 1,
%%\end{equation}
%%%
%%where $\vr_{o,{\rm ini}}$ and $\vv_o$ are a constant spatial coordinate and a constant velocity of the observer respectively.
The metric up to the first order of perturbations becomes
\begin{align}
\label{coor}
ds^2 &= -\left(1-2\sum_J\frac{GM_J}{|\vr-\vr_J|} -\frac{1}{3}\Lambda r^2\right) dt^2
\nonumber \\
&\quad + \left(1+2\sum_J\frac{GM_J}{|\vr-\vr_J|} -\frac{1}{6}\Lambda r^2\right)
d\vec{r}^2\ .
\end{align}
We note that the metric in this coordinate is spatially conformal flat, which reduces calculations of angles between two null geodesics to a simple one.
%Unlike the usual static coordinate of de Sitter spacetime, this isotropic coordinate can cover the spatial origin $\vr_P = \vec{0}$ also.
For the readers concerning the relation to the usual static coordinate of de Sitter spacetime, see appendix A.
%
%If the lensing object is a black hole located at the origin only, the spacetime becomes the Kottler spacetime. We can connect the present coordinate system to the one of Kottler by the coordinate transformation
%\begin{align}
% r_P &\simeq r - GM_{B} + \frac{1}{12}\Lambda r^3, \\
% M_B&: {\rm black\ hole\ mass} \nonumber
%\end{align}
%%
%in the weak field region. See appendix for the detail.

Once we construct the metric, we can rearrange the null geodesic equations to the following form using the derivatives with respect to the coordinate time
%we can derive a spatial part of null geodesic for a light ray parametrized by the coordinate time
\begin{align}
\label{eom}
\ddot{{\vec{r}}}_P &= 2[\vec{a}_P - 2(\vec{a}_P\cdot \vec{v}_P) \vec{v}_P], \\
\vec{a}_P&:= - \sum_J \frac{GM_J\vr_{PJ}}{r_{PJ}^3} + \frac{1}{12}\Lambda \vr_{P}, \nonumber \\
\vec{v}_P&:= \dot{\vec{r}}_P\ , \nonumber \\
\vr_{PJ}&:= \vr_P -\vr_J,\quad r_{PJ}:= |\vr_P -\vr_J|\ , \nonumber
\end{align}
where $\vec{r}_P$ is the position of the light on the null geodesic.

The (spatial) acceleration $\vec{a}_P$ consists of the usual Newtonian term and the cosmological constant term which gives a repulsive force if $\Lambda > 0$. Hereafter we use an abbreviation for a difference between two spatial points as $\vr_{AB}:= \vr_A - \vr_B$.
We solve the above equation perturbatively around a uniform linear motion as
\begin{align}
\label{vsol}
&\vec{v}_P= \vec{v}_{s} + \delta\vec{v}_P - 2(\delta\vec{v}_P\cdot\vec{v}_s)\vec{v}_s,
\\
\label{rsol}
&\vec{r}_P= \vec{r}_{s} + \vec{v}_s(t-t_s) +
\delta\vec{r}_P - 2(\delta\vec{r}_P\cdot\vec{v}_s)\vec{v}_s .
%&\vec{v}_s,\ \vr_{s}: {\rm constants}
\end{align}
$\vec{v}_s$ and $\vr_s$ are the initial velocity and the initial position of light at a light source which refer to the uniform linear motion. $t_s$ is the initial coordinate time when the light leaves the source.
$\delta\vec{v}_P$ and $\delta\vr_P$ are the first-order corrections from the leading uniform linear motion.
With the use of the null condition for the four-velocity of light, up to the first order of perturbations,\footnote{
$\vec{v}_P$ is an apparent velocity of light so that it can be slower or faster than a locally measured light speed $c(=1)$ depending on the signature of $\Lambda$.}
\begin{equation}
\label{onshell}
v_P^2 = 1 - \sum_J\frac{4GM_J}{r_{PJ}} -\frac{1}{6}\Lambda r_{P}^2 \ ,
\end{equation}
we can write the corrections explicitly as
\begin{align}
\delta\vv_P&= \sum_J\frac{2GM_J}{(\vv_s\times\vr_s)^2}
\left[\frac{(\vr_{sJ}\cdot\vr_{PJ})\vv_s - (\vv_s\cdot\vr_{PJ})\vr_{sJ}}{r_{PJ}}
\right. \nonumber \\
&\quad \left.-\frac{r_{sJ}^2\vv_s-(\vv_s\cdot\vr_{sJ})\vr_{sJ}}{r_{sJ}}\right]
\nonumber \\
%&\quad +\frac{1}{12}\Lambda[2\vr_{so}(t-t_s)+\vv_s(t-t_s)^2] \ , \\
&\quad +\frac{1}{12}\Lambda[2\vr_{s}(t-t_s)+\vv_s(t-t_s)^2] \ , \\
\delta\vr_P&= \sum_J\frac{2GM_J}{(\vv_s\times\vr_s)^2}
\left[((\vv_{s}\cdot\vr_{sJ})\vv_s - \vr_{sJ})(r_{PJ}-r_{sJ})
\right. \nonumber \\
&\quad -\frac{r_{sJ}^2\vv_s-(\vv_s\cdot\vr_{sJ})\vr_{sJ}}{r_{sJ}}(t-t_s)
\nonumber \\
&\quad \left.+((\vv_{s}\cdot\vr_{sJ})^2 - r_{sJ}^2)\vv_s
{\rm ln}\left|\frac{r_{sJ}+\vv_s\cdot\vr_{sJ}}{r_{PJ}+\vv_s\cdot\vr_{PJ}}
\right|\right] \nonumber \\
&\quad +\frac{1}{36}\Lambda[3\vr_{s}(t-t_s)^2+\vv_s(t-t_s)^3] \ , \\
%&\quad +\frac{1}{36}\Lambda[3\vr_{so}(t-t_s)^2+\vv_s(t-t_s)^3] \ , \\
\vv_s\times\vr_s& := (v_s^yz_s-v_s^zy_s,\ v^z_sx_s -v^x_sz_s,\ v^x_sy_s-v^y_sx_s)\ . \nonumber
\end{align}
This result corresponds to the textbook solution if $\Lambda = 0$.

For optical observations in the astrophysics, we are often interested into a light ray propagating from a light source to an observer. Using Eq.(\ref{rsol}), we derive the propagation time of light from the light source as
\begin{align}
t- t_s&= r_{Ps} - \sum_J2GM_J{\rm ln}\left|\frac{r_{sJ}+\vr_{Ps}\cdot\vr_{sJ}}
{r_{PJ}+\vr_{Ps}\cdot\vr_{PJ}}\right| \nonumber \\
&+\frac{1}{12}\Lambda\left(r_{Ps}r_{s}^2 +\frac{r_{Ps}^3}{3}
+(\vr_{s}\cdot\vr_{Ps})r_{Ps}\right) \ .
%&+\frac{1}{12}\Lambda\left(r_{Ps}r_{os}^2 +\frac{r_{Ps}^3}{3}
%-(\vr_{os}\cdot\vr_{Ps})r_{Ps}\right) \ .
\end{align}
Substituting this back into Eq.(\ref{rsol}) and identifying $\vr_P$ to the position of the observer $\vr_o$, we can express the initial velocity of light propagating from the light source to the observer as
\begin{align}
\label{vs}
\vv_s&= \frac{\vr_{os}}{r_{os}}\left(1 - \sum_J\frac{4GM_J}{r_{sJ}}\right) \nonumber \\
&+\sum_J\frac{2GM_Jr_{os}}{r_{oJ}r_{sJ}+\vr_{oJ}\cdot\vr_{sJ}}
\left(\frac{r_{oJ}+r_{sJ}}{r_{os}}\frac{\vr_{os}}{r_{os}}+\frac{\vr_{sJ}}{r_{sJ}}
\right) \nonumber \\
&-\frac{1}{12}\Lambda\left[ (r_s^2 -\vr_s\cdot\vr_{os})
\frac{\vr_{os}}{r_{os}} + r_{os}\vr_s \right] \ .
%&-\frac{1}{12}\Lambda r_{os}\vr_{os}\ .
\end{align}
Substituting this into Eq.(\ref{vsol}), we obtain the light velocity at the observer
\begin{align}
\label{ve}
\vv_P(t_s;t_o)&= \frac{\vr_{os}}{r_{os}}\left(1 - \sum_J\frac{4GM_J}{r_{oJ}}\right) \nonumber \\
&+\sum_J\frac{2GM_Jr_{os}}{r_{oJ}r_{sJ}+\vr_{oJ}\cdot\vr_{sJ}}
\left(\frac{r_{oJ}+r_{sJ}}{r_{os}}\frac{\vr_{os}}{r_{os}}-\frac{\vr_{oJ}}{r_{oJ}}
\right) \nonumber \\
&-\frac{1}{12}\Lambda\left[ (r_{os}^2 + r_s^2 +3\vr_s\cdot\vr_{os})
\frac{\vr_{os}}{r_{os}} - r_{os}\vr_s\right] \ ,
\end{align}
where $t_o$ is the coordinate time at the observer.
We can easily check that Eq.(\ref{vs}) and (\ref{ve}) satisfy the null condition (\ref{onshell}).

%他の座標系移りたかったら、0次成分含めて座標変換すればいいくらいのこと書いとくか？いらんか。相対論で座標変換で移ればいいなんて当たり前すぎる。CPTだって全部comoving gaugeとかで計算してるし。

In the observations, we receive the light which has the apparent velocity given by Eq.(\ref{ve}).
However, Eq.(\ref{ve}) depends on how to choose the coordinate system, so it cannot be an observable by itself. We will define angle differences in a coordinate independent way and connect them to the apparent velocity (\ref{ve}) in the next section.

%%=================================================================
\section{Angle differences}\label{3}

%We define an angle difference made by two null geodesics entering the observer.

%When we observe a single right ray, we cannot determine an incidence angle of the ray since there is no standard axis in the sky. To determine the angle, we usually use a reference system like the international celestial reference frame and measure angle differences between the light from the star and
%%We can measure, however, an angle difference between two light rays like a reference star and a target star.
%%Thus, as an observable, we define the angle difference made by two null geodesics entering the observer in a coordinate independent way.
In this section, as observables, we define coordinate invariant angle differences between two null geodesics which enter the observer, using a local rest frame of the observer.

We write four-momenta of null geodesics coming from light sources $s_i$ as $k^\mu_{s_i}$.
$\{s_i\}$ labels each light source.
All of the observables are defined and measured in the observer\rq{}s local rest frame.
So, we first introduce a tetrad which is a map to the local Lorentz frame as follows:
\begin{align}
{e^{\hat{0}}}_\mu &= (1/\sqrt{|g^{00}|},\ 0,\ 0,\ 0), \nonumber \\
{e^{\hat{1}}}_\mu &= (0,\ \sqrt{g_{11}},\ 0,\ 0), \nonumber \\
{e^{\hat{2}}}_\mu &= (0,\ 0,\ \sqrt{g_{22}},\ 0), \nonumber \\
{e^{\hat{3}}}_\mu &= (0,\ 0,\ 0,\ \sqrt{g_{33}}), \nonumber \\
%{\rm diag}(\sqrt{|g_{00}|},\, \sqrt{g_{11}},\, \sqrt{g_{22}},\, \sqrt{g_{33}}) \ ,
%{e^{\hat{\alpha}}}_{\mu}&= {\rm diag}(\sqrt{|g_{00}|},\, \sqrt{g_{xx}},\, \sqrt{g_{yy}},\, \sqrt{g_{zz}})\ ,\\
k^{\hat{\alpha}}_{s_i} &= {e^{\hat{\alpha}}}_{\mu}k^\mu_{s_i} \nonumber\ .
\end{align}
%
%The $\hat{0}$ component of this tetrad corresponds to a unit normal of a constant hypersurface with respect to the time.
%The tetrad maps a four velocity of a timelike observer $u^\mu$ to the local Lorentz frame as
%Given a four velocity vector of a time like observer in the coordinate basis as $u^\mu = u^0(1, \ v^i)$,
A four velocity vector of a timelike observer, $\bf{u}$, is represented in the coordinate basis $\{\partial_\mu\}$ and in the present tetrad basis $\{{\rm e}_{\hat{\alpha}}\}$ as
%\begin{equation}
% u^\mu = u^0(1,v^i)\ \rightarrow \ u^{\hat{\alpha}} = u^0(1/\sqrt{|g^{00}|},\ \sqrt{g_{ij}}v^j)\ .
%\end{equation}
%
\begin{align}
{\bf u}&= u^\mu\partial_\mu = u^{\hat{\alpha}}{\rm e}_{\hat{\alpha}} \ , \nonumber \\
u^\mu \partial_\mu &= u^0\partial_0+ u^0v^i\partial_i \ , \nonumber \\
u^{\hat{\alpha}}{\rm e}_{\hat{\alpha}}&= \frac{u^0}{\sqrt{|g^{00}|}}{\rm e}_{\hat{0}} + u^0\sqrt{g_{ij}}v^j{\rm e}_{\hat{i}}\ ,
% u^\mu{\rm e}_\mu = u^{\hat{\alpha}}{\rm e}_{\hat{\alpha}}\ , \nonumber \\
\end{align}
where $v^i$ is the apparent velocity of the observer.
The local rest frame of the observer is represented by another tetrad basis $\{{\rm e}_{\hat{\alpha}\rq{}}\}$ satisfied with ${\bf u} = {\rm e}_{\hat{0}\rq{}}$.
%In general, an observer is not always at rest, and we need to boost the system to the observer\rq{}s local rest frame where ${\bf u} = {\rm e}_{\hat{0}\rq{}}$.
%Such a Lorentz boost is given as
To transform the basis $\{{\rm e}_{\hat{\alpha}}\}$ to $\{{\rm e}_{\hat{\alpha}\rq{}}\}$, we need a Lorentz boost given as
\begin{align}
\label{boost}
\Lambda^{\hat{\alpha}}_{\ \hat{\beta}} &= \left(
\begin{array}{cc}
\hat{\gamma} & -\hat{\gamma} v^{\hat{j}} \\
-\hat{\gamma} v^{\hat{i}} & \delta_{\hat{i}\hat{j}}
+ (\hat{\gamma}-1)\frac{v^{\hat{i}}v^{\hat{j}}}{\hat{v}^2}
\end{array}
\right) \ , \\
\hat{v} &:= \sqrt{v^{\hat{k}}v^{\hat{k}}} , \quad \hat{\gamma}:= 1/\sqrt{1-\hat{v}^2} \ , \nonumber \\
u^{\hat{\alpha}\rq{}} &= \Lambda^{\hat{\alpha}}_{\ \hat{\beta}}u^{\hat{\beta}} \ , \nonumber
\end{align}
where the local three-velocity of the observer $v^{\hat{i}}$ relates to the apparent velocity of the observer as
\begin{equation}
v^{\hat{i}} = \sqrt{|g^{00}|g_{ij}}v^j\ .
\end{equation}
%
%Since we approximate the observer\rq{}s motion as in Eq.(\ref{ro}),
%up to the first order, we obtain
%\begin{equation}
% v^{\hat{i}} \simeq v^i_o \ .
%\end{equation}
%
%All of the observables are measured and defined in the boosted local inertial frame where the observer is at rest.
Using the boosted null momentum
\begin{align}
k^{\hat{i}\rq{}}_{s_i} &= \Lambda^{\hat{i}}_{\ \hat{\alpha}}k^{\hat{\alpha}}_{s_i} \nonumber \\
&= k^{\hat{i}}_{s_i} - \gamma v^{\hat{i}}k^{\hat{0}}_{s_i}
+(\gamma-1)\frac{k^{\hat{j}}_{s_i}v^{\hat{j}}}{\hat{v}^2}v^{\hat{i}} \ ,
\end{align}
we define
coordinate invariant angle differences (we call them declination and right ascension) made by spatial parts of two null momenta $k_{s_i}^{\hat{i}\rq{}}$ and
$k_{s_j}^{\hat{j}\rq{}}$ as
\begin{align}
&{\rm DEC}_{ij}:= {\rm arccos}\left(
\frac{k_{s_i}^{\hat{x}\rq{}}k_{s_j}^{\hat{x}\rq{}}
+\hat{l}_{s_i}\rq{}\hat{l}_{s_j}\rq{}}
%+\sqrt{{k_{s_i}^{\hat{y}\rq{}}}^2+{k_{s_i}^{\hat{z}\rq{}}}^2}
%\sqrt{{k_{s_j}^{\hat{y}\rq{}}}^2+{k_{s_j}^{\hat{z}\rq{}}}^2}}
{\sqrt{{k^{\hat{x}\rq{}}_{s_i}}^2+\hat{l}_{s_i}\rq{}^2}
\sqrt{{k^{\hat{x}\rq{}}_{s_j}}^2+\hat{l}_{s_j}\rq{}^2}}\right),
\nonumber \\
&{\rm RA}_{ij}:= {\rm arccos}\left(
\frac{k^{\hat{y}\rq{}}_{s_i}k^{\hat{y}\rq{}}_{s_j}
+k^{\hat{z}\rq{}}_{s_i}k^{\hat{z}\rq{}}_{s_j}}
{\hat{l}_{s_i}\rq{}\hat{l}_{s_j}\rq{}}\right), \nonumber \\
% {\sqrt{{k^{\hat{y}\rq{}}_{s_i}}^2+{k^{\hat{z}\rq{}}_{s_i}}^2}
%\sqrt{{k^{\hat{y}\rq{}}_{s_j}}^2+{k^{\hat{z}\rq{}}_{s_j}}^2}} \right) \ .
\label{AD}
&\hat{l}_{s_i}\rq{}:= \sqrt{{k^{\hat{y}\rq{}}_{s_i}}^2+{k^{\hat{z}\rq{}}_{s_i}}^2} \ .
\end{align}
${\rm DEC}_{ij}$ is the difference of the angles between $\hat{x}\rq{}$-axis and each null momentum
$k_{s_i}^{\hat{i}\rq{}},\ k_{s_j}^{\hat{j}\rq{}}$.
${\rm RA}_{ij}$ is the angle difference between $k_{s_i}^{\hat{i}\rq{}}$ and $k_{s_j}^{\hat{j}\rq{}}$
projected onto the $\hat{y}\rq{}\hat{z}\rq{}$-plane. See Fig.1.
\begin{figure}[h]
\begin{center} %センタリングする
\includegraphics[clip, pagebox=cropbox, height=5.0cm]{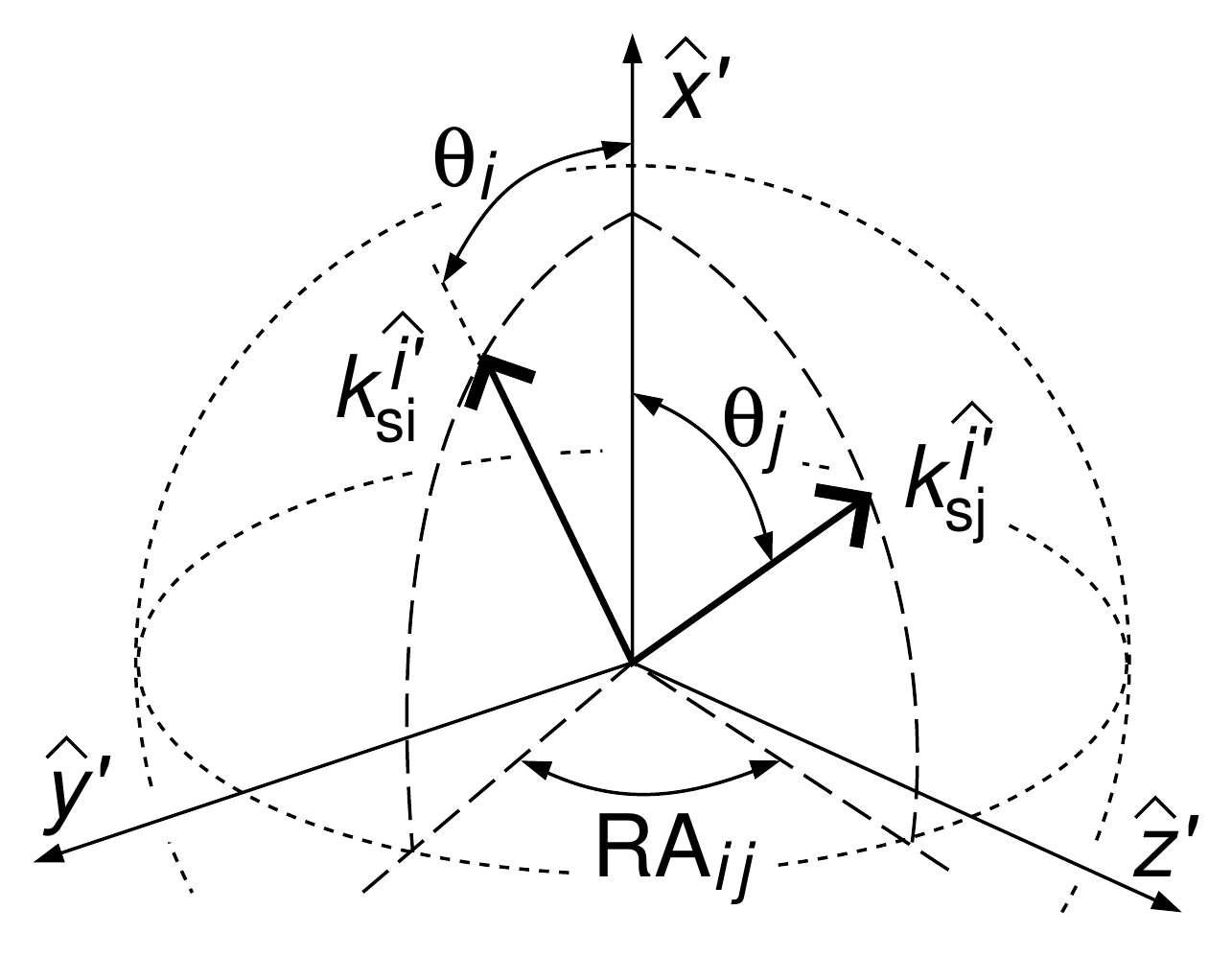}
\caption{The angle differences. The declination is defined as ${\rm DEC}_{ij}:= \theta_i-\theta_j$. The lights enter from below in this figure.}
\label{fig2} %ラベルをつけ図の参照を可能にする
\end{center}
\end{figure}
These definitions are conceptually the same as in \cite{angles}.
We may set $\hat{x}\rq{}$-axis and the other axes by the lights from some reference stars.
For the derivation of Eq.(\ref{AD}), see Appendix B.
We have constructed the angle differences by the scalars only, and hence they are invariant under coordinate transformations.
%We partially check the coordinate invariance of them by a calculation in another coordinate. See Appendix B.

Next, we evaluate the angle differences (\ref{AD}) perturbatively. We focus on the case that the angle differences are much smaller than unity and the observer moves slowly, $v^{\hat{i}}\ll c$.
%で、ブーストする前の系の言葉で書いていくよってこと。
Using a rotational transformation,
we can set the spatial axes satisfied with
\begin{equation}
|k^{\hat{z}\rq{}}_{s_i}| \gg |k^{\hat{x}\rq{}}_{s_i}|,\ |k^{\hat{y}\rq{}}_{s_i}| \ .
\end{equation}
Since we consider the small angle differences, we can obtain the above inequality for all the right sources $\{s_i\}$ entering our small viewing angle.
%for the light rays entering into a small viewing angle, and we define small quantities
We restrict ourselves to the slowly boosted case so that we get the same inequality for the null momenta before the boost also:
\begin{equation}
\label{22}
|k^{\hat{z}}_{s_i}| \gg |k^{\hat{x}}_{s_i}|,\ |k^{\hat{y}}_{s_i}| \quad {\rm for}\ \forall s_i\ .
\end{equation}
This implies $|k^{z}_{s_i}|\gg |k^{x}_{s_i}|,\ |k^{y}_{s_i}|$ in terms of the present coordinate system.
We can always set the spatial coordinates satisfied with this inequality, {\it without} changing the form of the metric
since the present coordinate system possesses rotational isometry.
Then, we define the small parameters as
\begin{equation}
\label{ratio}
\alpha_i:= \frac{k^{\hat{x}}_{s_i}}{k^{\hat{z}}_{s_i}},\quad
\beta_i:= \frac{k^{\hat{y}}_{s_i}}{k^{\hat{z}}_{s_i}}\ .
\end{equation}
%
%パラメーターの名前を変える。decやraとの混同を避けるため。
%
We use Eq.(\ref{ratio}) and the observer\rq{}s velocity $v^{\hat{i}}$ as small parameters for the perturbative expansion.
With the use of the null condition on a local Lorentz frame
\begin{equation}
|k^{\hat{0}}_{s_i}| = |k^{\hat{z}}_{s_i}|\sqrt{1+\alpha_i^2 + \beta_i^2}\ ,
\end{equation}
we can evaluate the small angle differences perturbatively as
\begin{align}
\label{dangle}
&{\rm DEC}_{ij} \nonumber \\
&\simeq \frac{k_{s_i}^{\hat{x}'}}{k_{s_i}^{\hat{z}'}}
- \frac{k_{s_j}^{\hat{x}\rq{}}}{k_{s_j}^{\hat{z}\rq{}}}
+O\Big(\hat{v}\cdot (\alpha_i^2,\beta_i^2),\
\hat{v}^2\cdot(\alpha_i,\beta_i)\Big) \nonumber \\
&= (\alpha_i - \alpha_j)(1+v^{\hat{z}}) \nonumber \\
&\quad+O\Big(\hat{v}\cdot (\alpha_i^2,\beta_i^2),\
\hat{v}^2\cdot(\alpha_i,\beta_i)\Big), \nonumber \\
&{\rm RA}_{ij} \nonumber \\
&\simeq \frac{k_{s_i}^{\hat{y}'}}{k_{s_i}^{\hat{z}'}}
- \frac{k_{s_j}^{\hat{y}\rq{}}}{k_{s_j}^{\hat{z}\rq{}}}
+O\Big(\hat{v}\cdot (\alpha_i^2,\beta_i^2),\
\hat{v}^2\cdot(\alpha_i,\beta_i)\Big) \nonumber \\
%\left(1+O(\hat{v}^2, {\rm RA}_i^2)\right) \nonumber \\
&= (\beta_i - \beta_j)(1+v^{\hat{z}}) \nonumber \\
&\quad +O\Big(\hat{v}\cdot (\alpha_i^2,\beta_i^2),\
\hat{v}^2\cdot(\alpha_i,\beta_i)\Big),
\end{align}
where we ignore the third and higher order terms of ($\alpha_i,\,\beta_i$).
%up to $O(\alpha_i\cdot \hat{v},\,{\rm RA}_i\cdot \hat{v})$.
Up to this order, the effect of boost corresponds to that of a simple Galilei transformation.

In the present coordinate system (\ref{coor}), due to the spatially conformal flatness, we can reduce Eq.(\ref{ratio}) to
\begin{align}
\label{ratiov}
\alpha_i
= \frac{v^x_P(t_{s_i};t_o)}{v^z_P(t_{s_i};t_o)},\quad
\beta_i
= \frac{v^y_P(t_{s_i};t_o)}{v^z_P(t_{s_i};t_o)}\ .
\end{align}
where $v_P^i(t_{s_i};t_o)$ is given by Eq.(\ref{ve}).
Expanding Eq.(\ref{ratiov}) around the leading solutions obtained for the flat spacetime, we get
\begin{align}
\label{deci}
\alpha_i
&= \alpha_i^{(0)} -\frac{1}{12}\Lambda r_{os_i}^2
\frac{z_{s_i}x_o - x_{s_i}z_o}{(z_o-z_{s_i})^2} \nonumber \\
&-\sum_J \frac{2GM_J}{r_{oJ}}\left[
\frac{(z_o-z_J)(x_{s_i}-x_J)}{r_{oJ}r_{s_iJ}+\vr_{oJ}\cdot\vr_{s_iJ}}\right. \nonumber \\
&\quad -\left.\frac{(x_o-x_J)(z_{s_i}-z_J)}{r_{oJ}r_{s_iJ}+\vr_{oJ}\cdot\vr_{s_iJ}}\right] \nonumber \\
%&+\frac{1}{12}\Lambda(x_{s_i}z_o - z_{s_i}x_o) \nonumber \\
&+O\left( \frac{GM_J}{r_{oJ}}
\cdot ({\alpha_i^{(0)}}^2, {\beta_i^{(0)}}^2)\right) \\
\label{rai}
\beta_i
&= \beta_i^{(0)}-\frac{1}{12}\Lambda r_{os_i}^2
\frac{z_{s_i}y_o - y_{s_i}z_o}{(z_o-z_{s_i})^2} \nonumber \\
&-\sum_J \frac{2GM_J}{r_{oJ}}\left[
\frac{(z_o-z_J)(y_{s_i}-y_J)}{r_{oJ}r_{s_iJ}+\vr_{oJ}\cdot\vr_{s_iJ}}\right. \nonumber \\
&\quad-\left.\frac{(y_o -y_J)(z_{s_i}-z_J)}{r_{oJ}r_{s_iJ}+\vr_{oJ}\cdot\vr_{s_iJ}}\right] \nonumber \\
%&+\frac{1}{12}\Lambda(y_{s_i}z_o - z_{s_i}y_o) \nonumber \\
&+O\left( \frac{GM_J}{r_{oJ}}
\cdot ({\alpha_i^{(0)}}^2, {\beta_i^{(0)}}^2)\right) \ ,
\end{align}
where
\begin{equation}
\alpha_i^{(0)}:= \frac{x_o - x_{s_i}}{z_o- z_{s_i}},\quad
\beta_i^{(0)}:= \frac{y_o - y_{s_i}}{z_o- z_{s_i}}
\end{equation}
are small parameters for the case of flat spacetime as the same as Eq.(\ref{ratio}).
Substituting them to Eq.(\ref{dangle}), we finally obtain a perturbative expansion of the small angle differences by means of the present coordinate with clarifying the magnitude of the approximation errors.
\section{A Practical example}\label{4}

In this section, we evaluate the effect of the lensing objects on the small angle differences by using the results of the previous section for a practical example.

First, we identify where and who the observer is.
We consider the observer standing at the spatial origin $\vr_o=\vec{0}$, who watches the isotropic cosmic space except for the lensing objects.
This observer corresponds to an observer who observes the homogeneous and isotropic cosmic space on the flat slicing of the de Sitter spacetime.
In reality, we do not necessarily see the (local) isotropic cosmic space\footnote{
In fact, we move slowly against CMB and do not observe the isotropic cosmic space. The same is true for the local cosmic space, but the relative velocity of us to the local cosmic space does not need to correspond to the relative velocity of us to CMB.},
so that the angle differences observed by us and the observer at $\vr_o=\vec{0}$ may deviate each other.
However, if we move very slowly,
%and the cosmological constant is sufficiently small,
we can ignore its effect on the angle differences.
Here, we evaluate the angle differences regarding ourselves as the rest observer at the spatial origin.

Next, we consider the lensing objects.
We give the solar system and Sagittarius A*(Sgr A*), a supermassive black hole candidate at our galactic center, as the lensing objects, and consider stars or flares near around Sgr A*, like S-stars, as the light sources.
In the sum of lensing objects, we denote the solar system objects by the index $I$ and Sgr A* by the index $B$:
$\sum_J = \sum_{I, B}$.
%We capture Sgr A* and all the light sources around it in the field of view and take $z$-axis to the line of sight at a certain time.
We set the $z$-axis as the null momenta from the light sources around Sgr A* are satisfied with $|k^z_{s_i}|\gg |k^x_{s_i}|,\ |k^y_{s_i}|$ and thus Eq.(\ref{22}) also.
This $z$-axis is almost the same as the line of sight to the light sources at a certain time.
%これは、ある時刻におけるそれらの光源の視線方向にｚじくをつけるのとほぼ同義。
Also, we set the direction of $z$-axis as $z_o-z_{s_i}> 0$. Since Sgr A* and the light sources are very far from the observer and the solar system, we approximately get
\begin{align}
\label{33}
\frac{x_o-x_B}{z_o-z_B}\sim \alpha^{(0)}_{i},\ \frac{y_o-y_B}{z_o-z_B}\sim \beta^{(0)}_{i},
\ \vr_{s_iI}\sim \vr_{s_io},
\end{align}
which leads to
\begin{align}
\label{34}
r_{oB}&\simeq (z_o-z_B)(1+O({\alpha_i^{(0)}}^2,{\beta_i^{(0)}}^2)), \nonumber \\
r_{s_iI}&\simeq (z_I - z_{s_i})(1+O({\alpha_i^{(0)}}^2,{\beta_i^{(0)}}^2)).
\end{align}
$(x_B, y_B, z_B)$ denotes the spatial coordinates of Sgr A*.
Using Eq.(\ref{33}) and (\ref{34}), we can reduce Eq.(\ref{deci}) and (\ref{rai}) to
%\begin{align}
% {\rm DEC}_{s_i}&\simeq {\rm DEC}_{s_i}^{(0)} \nonumber \\
% &+\sum_I\frac{2GM_I}{r_{oI}}\frac{x_o-x_I}{r_{oI}(1+\vec{n}_{oI}\cdot\vec{n}_{s_iI})} \nonumber \\
% &+\frac{2GM_B}{r_{oB}}\frac{x_{s_i}-x_B}{r_{s_iB}(1+\vec{n}_{oB}\cdot\vec{n}_{s_iB})} \nonumber \\
% &-\frac{1}{12}\Lambda (x_{s_i}z_E-z_{s_i}x_E) \nonumber \\
% &+O\Big(GM_J/r_{EJ}\cdot ({\rm DEC}_{s_i}^{(0)},{\rm RA}_{s_i}^{(0)}),\ \nonumber \\
% &\qquad \Lambda r_{Es_i}^2\cdot ({{\rm DEC}_{s_i}^{(0)}}^2,{{\rm RA}_{s_i}^{(0)}}^2) \Big) \\
% \vec{n}_{AB}&:= \frac{\vr_A - \vr_B}{r_{AB}}
%\end{align}
%%
%
%\begin{align}
% \vec{n}_{EI}\cdot\vec{n}_{s_iI}&\simeq -\frac{z_E-z_I}{r_{EI}}
% + O({\rm DEC}_{s_i}^{(0)},{\rm RA}_{s_i}^{(0)}) \nonumber \\
% \vec{n}_{EB}\cdot\vec{n}_{s_iB}&\simeq \frac{z_{s_i}-z_B}{r_{s_iB}}
% + O({\rm DEC}_{s_i}^{(0)},{\rm RA}_{s_i}^{(0)})
%\end{align}
%%
%
\begin{align}
\alpha_i&\simeq \alpha_i^{(0)}-\sum_I\frac{2GM_I}{r_{oI}}\frac{x_o-x_I}{r_{oI}+z_I-z_o} \nonumber \\
&-\frac{2GM_B}{r_{oB}}\frac{x_{s_i}-x_B}{r_{s_iB}+z_{s_i}-z_B} \nonumber \\
%&-\frac{1}{12}\Lambda (x_{s_i}z_E-z_{s_i}x_E) \nonumber \\
&+O\left(\frac{GM_J}{r_{oJ}}\cdot (\alpha_i^{(0)},\beta_i^{(0)})\right), \\
\beta_i&\simeq \beta_i^{(0)} -\sum_I\frac{2GM_I}{r_{oI}}\frac{y_o-y_I}{r_{oI}+z_I-z_o} \nonumber \\
&-\frac{2GM_B}{r_{oB}}\frac{y_{s_i}-y_B}{r_{s_iB}+z_{s_i}-z_B} \nonumber \\
%&-\frac{1}{12}\Lambda (x_{s_i}z_E-z_{s_i}x_E) \nonumber \\
&+O\left(\frac{GM_J}{r_{oJ}}\cdot (\alpha_i^{(0)},\beta_i^{(0)})\right)\ .
%&\qquad \Lambda r_{os_i}^2\cdot ({{\rm DEC}_{s_i}^{(0)}}^2,{{\rm RA}_{s_i}^{(0)}}^2) \Big) \ .
\end{align}
The cosmological constant term vanishes for the rest observer at the spatial origin.
For the sun, $GM_\odot/r_{o\odot}\sim 5$ miliarcsecond (mas), and for Sgr A*, $GM_B/r_{oB}\sim 10$ microarcsecond ($\mu$as) using estimated values for their masses and distances. The leading terms $\alpha_i^{(0)}$ and $\beta_i^{(0)}$ should be larger than these values not to break down the perturbative expansion.
In the present approximation, the deflection of light by the solar system {\it does not} depend on the positions of each sources $\vec{r}_{s_i}$ while the deflection by Sgr A* does. When we derive the angle differences, therefore, the deflection by the solar system cancels out and only the deflection by Sgr A* remains:
\begin{align}
\label{resu}
{\rm DEC}_{ij}
&= \alpha_i^{(0)} - \alpha_{j}^{(0)} \nonumber \\
&-\frac{2GM_B}{r_{oB}}\left(
\frac{x_{s_i}-x_B}{r_{s_iB}+z_{s_i}-z_B}-\frac{x_{s_j}-x_B}{r_{s_jB}+z_{s_j}-z_B}
\right) \nonumber \\
&+ O\left(
\frac{GM_J}{r_{oJ}}\cdot (\alpha_i^{(0)},\beta_i^{(0)})\right), \nonumber \\
{\rm RA}_{ij}
&= \beta_i^{(0)} - \beta_{j}^{(0)} \nonumber \\
&-\frac{2GM_B}{r_{oB}}\left(
\frac{y_{s_i}-y_B}{r_{s_iB}+z_{s_i}-z_B}-\frac{y_{s_j}-y_B}{r_{s_jB}+z_{s_j}-z_B}
\right) \nonumber \\
&+ O\left(
\frac{GM_J}{r_{oJ}}\cdot (\alpha_i^{(0)},\beta_i^{(0)})\right) \ .
\end{align}
%\begin{align}
% &{\rm DEC}_{ij} \nonumber \\
% &= \left(\alpha_i^{(0)} - {\rm DEC}_{j}^{(0)}\right)(1+v^{z}_o) \nonumber \\
% &-\frac{2GM_B}{r_{oB}}\left(
% \frac{x_{s_i}-x_B}{r_{s_iB}+z_{s_i}-z_B}-\frac{x_{s_j}-x_B}{r_{s_jB}+z_{s_j}-z_B}
% \right) \nonumber \\
% %+&\frac{1}{12}\Lambda \left[ (x_{s_1}z_E-z_{s_1}x_E) - (x_{s_2}z_E-z_{s_2}x_E)
% % \right] \nonumber \\
% &+ O\left(
% \frac{GM_J}{r_{oJ}}\cdot (\alpha_i^{(0)},\beta_i^{(0)}),
% \frac{GM_B}{r_{oB}}\cdot \hat{v}, \right.\ \nonumber \\
% %&\quad \Lambda r_{os_i}^2\cdot
% % (v^{\hat{z}}, {\alpha_i^{(0)}}^2,{\beta_i^{(0)}}^2) , \nonumber \\
% &\qquad \hat{v}\cdot (\alpha_i^2,\beta_i^2),\
% \hat{v}^2\cdot(\alpha_i,\beta_i)\Big), \\
% % &\quad {\rm DEC}_{12}^3\Big) \nonumber \\
% &{\rm RA}_{ij} \nonumber \\
% &= \left(\beta_i^{(0)} - {\rm RA}_{j}^{(0)}\right)(1+v^{z}_o) \nonumber \\
% &-\frac{2GM_B}{r_{oB}}\left(
% \frac{y_{s_i}-y_B}{r_{s_iB}+z_{s_i}-z_B}-\frac{y_{s_j}-y_B}{r_{s_jB}+z_{s_j}-z_B}
% \right) \nonumber \\
% %+&\frac{1}{12}\Lambda \left[ (x_{s_1}z_E-z_{s_1}x_E) - (x_{s_2}z_E-z_{s_2}x_E)
% % \right] \nonumber \\
% &+ O\left(
% \frac{GM_J}{r_{oJ}}\cdot (\alpha_i^{(0)},\beta_i^{(0)}),
% \frac{GM_B}{r_{oB}}\cdot \hat{v}, \right.\ \nonumber \\
% %&\quad \Lambda r_{os_i}^2\cdot
% % (v^{\hat{z}}, {\alpha_i^{(0)}}^2,{\beta_i^{(0)}}^2) , \nonumber \\
% &\qquad \hat{v}\cdot (\alpha_i^2,\beta_i^2),\
% \hat{v}^2\cdot(\alpha_i,\beta_i)\Big)\ .
%\end{align}
%
We consider the rest observer, so we set $v^{\hat{i}}=0$.
The deflection by the solar system exists certainly and can be the largest relativistic correction. We can regard it, however, as if it does not exist because of the canceling out in the angle differences.

In the case that a flare and S0-2, a star orbiting around Sgr A*, are the light sources,
we obtain $\sim100$ mas for the angle differences between them with $0.3$ mas errors by the latest observation \cite{Abuter:2020dou}.
\begin{figure}[h]
\begin{center} %センタリングする
\includegraphics[clip,pagebox=cropbox, height=5.0cm]{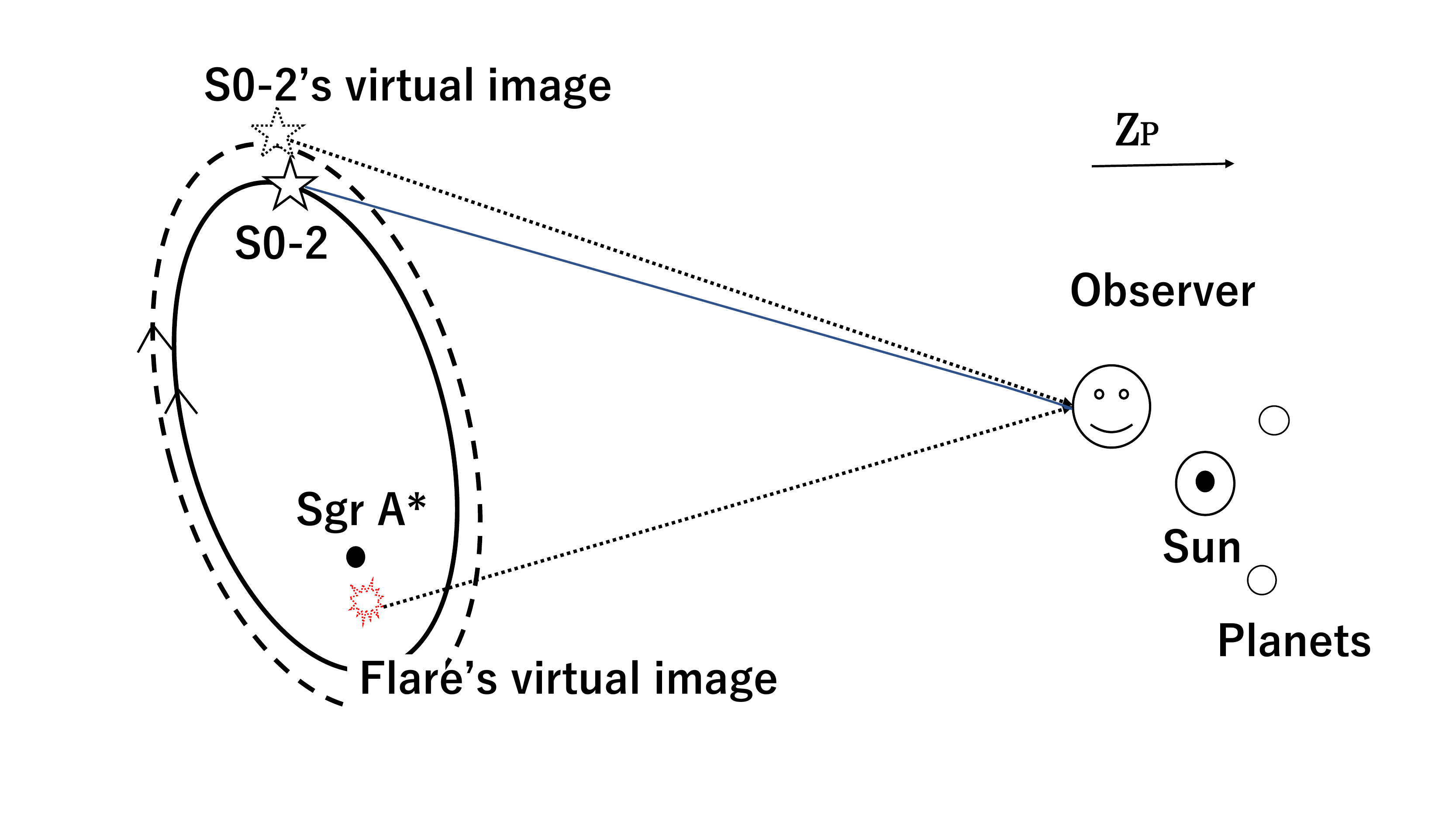}
\caption{The schematic picture of S0-2, the (virtual image of) flare near Sgr A* and the solar system. The spatial distances between the objects are extremely scaled up or down. S0-2 orbits around Sgr A*(real oval line) and the light from S0-2 (real blue line) enters the observer.
%who moves slowly relative to the galactic center.
The incident lights from S0-2 and the flare show us the virtual images of them. The angle between the real image and the virtual image consists of $\sim 10$ $\mu$as by Sgr A* and $\sim 5$ mas by the solar system, respectively. We do not need to take into account the deflections by the solar system for the lights from S0-2 and the flare since the deflections cancel out between the two lights.}
\label{fig1} %ラベルをつけ図の参照を可能にする
\end{center}
\end{figure}
%%The relative velocity in $z$-direction of the earth to the galactic center is estimated as $\sim10^{-4}$, so the effect of boost to the angle differences is $100\ {\rm mas}\times 10^{-4}\sim 10$ $\mu$as, comparable to the bending by Sgr A*. The error from the formula is suppressed around to $0.1$ $\mu$as.
Using the result (\ref{resu}), we find that the deflection by Sgr A* contributes about $10\ \mu$as to the angle differences, which would be detectable
in the next-generation observational instrument.
The approximation errors are suppressed around $0.1\ \mu$as at most.
Thus, we now become able to verify the general relativistic effect to the light ray more precisely by comparing the near-future observational data to our formula derived here.
\footnote{
In practice, the flare does not shine every time. We derive a point in the sky where the virtual image of flare shined in the past by using reference stars, and we measure the angle differences between that point and the virtual image of S0-2. If we do not concern about the true place of the flare, it is enough to take into account the bending for the light leaving from S0-2 only, not for the flare. That is, we express the angle differences between S0-2 and the flare by the true place of S0-2 and the virtual image's place of flare.
Even if we do not identify the true place of flare, anyway, the angle differences get $\sim10$
$\mu$as contribution from the deflection by Sgr A*.} %and the boost for the light leaving from S0-2.}
We note that our expression (\ref{resu}) can be easily applied for the model fitting to the actual observational data
since they concretely consist of the spatial points of the light sources and the lensing objects parametrized by the coordinate time.
For the observation of S0-2 and Sgr A*, the interval of proper time of us is almost the same as the interval of coordinate time, so that we can directly compare the theoretical expression (\ref{resu}) parametrized by the coordinate time to the observational data.
We evaluate the same angle differences in another coordinate system and confirm that the angle differences are invariant under the coordinate transformation at least up to the present order of approximation. See Appendix C.

Why does not the cosmological constant appear in the angle differences here?
This is because the angle differences are constructed by the tetrad and the null momenta at the spatial origin only,
%This is because we defined the angle differences by the tetrad and the null momenta at the spatial origin only,
where the effect of the cosmological constant vanishes from the metric tensor.
Thus, though the lights travel a long distance, the rest observer at the spatial origin does not explicitly observe the effect of the cosmological constant.
%%
%%だから手元のベクトル見てても宇宙項は入らない。けど、星の運動には効くはずだ。だから、光の初期条件には宇宙項の影響が乗ってくる的に。モニター観測で見ることができる。
%%without comparing to any physical quantities on other points.
%%We adopted the coordinate system, in which the effect of the cosmological constant vanishes at the observer standing at the spatial origin,
%%and the effect on the angle differences vanishes also. わかりにくい。
%%The light propagating from the light source to the observer has its momentum as a part of its Cauchy data.
%%The observer can create a light that has the same Cauchy data as its initial condition.
%%We cannot distinguish those two lights since they are completely the same physical quantities.
%%Thus, though the light travels a long distance, the rest observer at the spatial origin does not explicitly observe the effect of the cosmological constant.
%%%We could justify this argument by considering a light
%%%propagating from the observer to the light source. Since the equation of motion (\ref{eom}) possesses the time reversibility, the apparent velocity of the light at the observer is described by Eq.(\ref{ve}) with changing the signature only. Naturally, the velocity of light leaving at the observer should not take any effect of the cosmic expansion and so is the velocity of light arriving at the observer.
%%This reflects that momentum is tangent vector on each point.
%%%ソースから手元まで旅した光と全く同じエナジー、向きを初期条件としてもつ光をΛの影響がある時空点を旅していないにもかかわらず、手元で作ることができる。手元まで跳んできた光と手元で作った光を区別する物理量は何もない。だからオブザーバーの位置でΛの影響がないならどこから飛んできた光であろうがΛの影響はモメンタムに入らない。よって角度にも入らない。
%%This reflects that the cosmic expansion is measured by each observer.
%%
We note, however, that the cosmological constant affects the motions of the light sources.
For the null momenta, the positions of the light sources are a part of their initial conditions,
so that the cosmological constant is {\it implicitly} encoded into the angle differences through the positions of the light sources $\vec{r}_{s_i}$.
If we monitor the motions of the light sources such as S0-2 continuously, we could find the effect of the cosmological constant from the time evolution of the positions of the light sources.

How will the situation change for a slowly moving observer around the spatial origin?
If we regard the position of the observer is just a small variation from the spatial origin as
$\vr_{o}\simeq \vec{v}_{o} \delta t$ in the small-time interval $\delta t$ with the small velocity $\vv_o$,
the deviation of the angle differences remains around $\Lambda r_{s_io}|\vec{v}_o|\delta t$.
This variation can be ignorable in the observation if the cosmological constant and the velocity of the observer are sufficiently small in the small time interval.

%============================================================
\section{Summary}\label{5}

We have derived the spatially conformal flat metric perturbatively which describes the de Sitter spacetime with multi-lensing objects.
We solved the null geodesic on that spacetime perturbatively around the uniform linear motion, parametrized by the coordinate time.
Using the two world points, the points of the light source and the observer, we expressed the apparent velocities of the light at the light source and at the observer.

Then, we have defined the angle differences between two light rays
entering the observer in the coordinate invariant way.
The angle differences are constructed by the scalars defined in the observer's local rest frame only so that they are coordinate invariant and hence can be observables.
We reduced the coordinate invariant formula for the small angle differences case.
For the small angle differences, due to the spatially conformal flatness of the present coordinate system, we can express them easily by means of the apparent velocities of the light at the observer.

Using the coordinate invariant formula, as a practical example,
we evaluated the angle differences between the light rays from the star S0-2 and the flare around Sgr A*. We considered Sgr A* and the solar system as the lensing objects.
The lights are deflected by both the solar system and Sgr A*.
Up to the first order of the relativistic corrections, however, we find that the deflections by the solar system cancel out and only the deflections by Sgr A* remain in the small angle differences. We have shown that the deflections by Sgr A* can be about $10\ \mu$as, which would be detectable in the near-future observation. Also, we confirmed the coordinate invariance of the angle differences calculating them in another coordinate system.

On the other hand, we find that the cosmological constant does not appear in the angle differences explicitly for the rest observer who sees the isotropic cosmic space. We defined the angle differences by the null momenta and the tetrad at the observer only, and in the coordinate we use, the effect of the cosmological constant at the observer vanishes.
For the case that the observer moves slowly, the effect on the angle differences is to be proportional to the cosmological constant and the velocity of the observer in the small time interval. If those quantities are sufficiently small, we can still ignore the effect of the cosmological constant for the angle differences.
We emphasize, however, that the effect of the cosmological constant is implicitly taken
into account in the motions of the light sources. Thus, if we can express the motions of the light sources with the cosmological constant, we may determine the effect of the cosmological constant comparing our formulae to the observational data of the angle differences. Monitoring the motions of the light sources such as S0-2 continuously, we expect to give some upper limit to the local cosmological constant.

%では、他にΛの効果を見るのに適した量はないか？
%geoにΛの影響がある点で定義されている物理量を
%Λを見ようと思ったら、空間距離の差が大きい2点で定義されている物理量、レッドシフトとかを比べるのは当たり前。
%他には、Λは章動に載ってくるという研究もあり（れｆ引く）。そういった

\acknowledgements
R.S. and H.S. were supported by JSPS KAKENHI, Grant-in-Aid for
Scientific Research (B) 19H01900.

%===========================================================
\appendix
\section{Static slice to isotropic slice at the first order of perturbation}\label{A}

We consider a coordinate transformation from the static coordinate of de Sitter spacetime denoted by $(t, \vec{r}_S)$, to the isotropic coordinate denoted by $(t, \vec{r})$ up to the first order of perturbation. For simplicity, we ignore all the lensing objects.
%\begin{align}
% ds^2&=-\left(1-\frac{2GM_B}{r_S} -\frac{1}{3}\Lambda r_S^2\right) dt^2
% \nonumber \\
% &\quad + \frac{dr^2}{1-\frac{2GM_B}{r} -\frac{1}{3}\Lambda r^2}
% + r^2d\Omega_2 \nonumber \\
% &=-\left(1-\frac{2GM_B}{r} -\frac{1}{3}\Lambda r^2\right) dt^2
% \nonumber \\
% &\quad + \frac{(dr/dr_P)^2}{1-\frac{2GM_B}{r} -\frac{1}{3}\Lambda r^2}dr_P^2 \nonumber \\
% &\quad + \frac{r^2}{r_P^2}\left(r_P^2d\Omega_2\right)
%\end{align}
%
First, we rewrite the static coordinate of de Sitter spacetime as follows:
\begin{align}
ds^2&=-\left(1-\frac{1}{3}\Lambda r_S^2\right) dt^2
+ \frac{dr_S^2}{1-\frac{1}{3}\Lambda r_S^2}
+ r_S^2d\Omega_2 \nonumber \\
&=-\left(1-\frac{1}{3}\Lambda r_S^2\right) dt^2
+ \frac{(dr_S/dr)^2}{1-\frac{1}{3}\Lambda r_S^2}dr^2
\nonumber \\
&\quad + \frac{r_S^2}{r^2}\left(r^2d\Omega_2\right) \ ,
\end{align}
where $d\Omega_2$ is the infinitesimal 2-dimensional solid angle.
Transforming to the isotropic one, we need a relation
\begin{align}
\frac{dr}{r} &= \frac{dr_S}{r_S\left(1-\frac{1}{3}\Lambda r_S^2\right)^{1/2} } \nonumber \\
&\simeq \left(\frac{1}{r_S} +\frac{1}{6}\Lambda r_S\right) dr_S \ .
\end{align}
In the second line, we expanded the denominator treating the $\Lambda$ term as a small quantity.
Integrating the both sides, without the integration constant, we obtain
\begin{align}
r&\simeq r_S +\frac{1}{12}\Lambda r_S^3 ,\\
%\frac{r^2}{r_P^2}&\simeq 1+\frac{2GM_B}{r_P} -\frac{1}{6}\Lambda r_P^2 \nonumber \\
\label{sdS}
ds^2 &= -\left(1-\frac{1}{3}\Lambda r^2\right) dt^2
\nonumber \\
&\quad + \left(1 -\frac{1}{6}\Lambda r^2\right)(dr^2+
r^2d\Omega_2)\ .
\end{align}
%
%(Including the integration constant just corresponds to a spatial translation.)
%This isotropic coordinate can cover the spatial origin $\vec{r}_P=\vec{0}$ where the static coordinate cannot cover.
The isotropic coordinate is a good approximation around the spatial origin as long as the $\Lambda$ term is smaller than unity.
If we want to include the lensing objects, we just need to add linear perturbations to the above metric and solve the perturbed Einstein equation as in Eq.(\ref{Eeq}).

%========================================================
\section{Derivation of the angle differences}\label{B}

We derive the angle differences (\ref{AD}).
First, we derive the right ascension ${\rm RA}_{ij}$.
An angle $\Theta_{ij}$ between two momenta $k_{s_i}^{\hat{i}\rq{}}$ and $k_{s_j}^{\hat{i}\rq{}}$ is given as
\begin{equation}
\Theta_{ij} = {\rm arccos}\left(\frac{k_{s_i}^{\hat{i}\rq{}}{k_{s_j}^{\hat{i}\rq{}}}}
{|k_{s_i}^{\hat{j}\rq{}}||k_{s_j}^{\hat{k}\rq{}}|}\right)
\end{equation}
We define ${\rm RA}_{ij}$ as a projection of $\Theta_{ij}$ to $\hat{y}\rq{}\hat{z}\rq{}$-plane.
That is achived by taking an angle between 2-dimensional momenta projected to $\hat{y}\rq{}\hat{z}\rq{}$-plane.
Denoting ${\hat{l}^{a\rq{}}}_{s_i}=(k_{s_i}^{\hat{y}\rq{}},\,k_{s_i}^{\hat{z}\rq{}})$,
we obtain
\begin{align}
{\rm RA}_{ij}&= {\rm arccos}\left( \frac{{\hat{l}^{a\rq{}}}_{s_i}{\hat{l}^{a\rq{}}}_{s_j}}
{|{\hat{l}^{b\rq{}}}_{s_i}||{\hat{l}^{c\rq{}}}_{s_j}|}\right) \nonumber \\
&={\rm arccos}\left(
\frac{k^{\hat{y}\rq{}}_{s_i}k^{\hat{y}\rq{}}_{s_j}
+k^{\hat{z}\rq{}}_{s_i}k^{\hat{z}\rq{}}_{s_j}}
{\hat{l}_{s_i}\rq{}\hat{l}_{s_j}\rq{}}\right) \ , \nonumber \\
\hat{l}_{s_i}\rq{} &= \sqrt{{k^{\hat{y}\rq{}}_{s_i}}^2+{k^{\hat{z}\rq{}}_{s_i}}^2}\ .
\end{align}
Next, we derive the declination ${\rm DEC}_{ij}$. We denote an angle between momentum $k_{s_i}^{\hat{i}\rq{}}$ and $\hat{x}\rq{}$-axis as $\theta_i$, which gives
\begin{align}
&{\rm cos}\theta_i = \frac{k^{\hat{x}\rq{}}_{s_i}}{|k^{\hat{i}\rq{}}_{s_i}|},\quad
{\rm sin}\theta_i = \frac{\hat{l}_{s_i}\rq{}}{|k^{\hat{i}\rq{}}_{s_i}|}\ .%, \nonumber \\
%&0\leq \theta_i \leq \pi \ .
\end{align}
We define ${\rm DEC}_{ij}$ as $\theta_i -\theta_j$ and thus we obtain
\begin{align}
{\rm DEC}_{ij}&= {\rm arccos}({\rm cos}(\theta_i -\theta_j)) \nonumber \\
&= {\rm arccos}\left(
\frac{k_{s_i}^{\hat{x}\rq{}}k_{s_j}^{\hat{x}\rq{}}
+\hat{l}_{s_i}\rq{}\hat{l}_{s_j}\rq{}}
%+\sqrt{{k_{s_i}^{\hat{y}\rq{}}}^2+{k_{s_i}^{\hat{z}\rq{}}}^2}
%\sqrt{{k_{s_j}^{\hat{y}\rq{}}}^2+{k_{s_j}^{\hat{z}\rq{}}}^2}}
{\sqrt{{k^{\hat{x}\rq{}}_{s_i}}^2+\hat{l}_{s_i}\rq{}^2}
\sqrt{{k^{\hat{x}\rq{}}_{s_j}}^2+\hat{l}_{s_j}\rq{}^2}}\right) \ .
\end{align}
%

%====================================================
\section{Coordinate invariance of the angle differences}\label{C}

%%We considered the rest observer at the spatial origin in Sec.\ref{4}.
We consider a coordinate transformation by which Sgr A* becomes to stand at the spatial origin in the new coordinate system.
We assume that the spatial part of the world line for Sgr A* in the coordinate (\ref{coor}) is given as
\begin{align}
&\vr_B = \vr_{B,{\rm ini}} + \vv_B t,\quad |\vv_B|\ll c, \nonumber \\
&\vr_{B,{\rm ini}},\ \vv_B : {\rm constants},
\end{align}
and transform it to $\vec{\tilde{r}}_B=\vec{0}$ in the new coordinate $(\tilde{t}, \vec{\tilde{r}})$.
We note that in the coordinate system (\ref{coor}), Sgr A* and the other lensing objects move slowly though we can ignore their velocities in the metric and the energy-momentum tensor as the first-order approximation. They move, however, against the rest observer considered in Sec.\ref{4}.
By the coordinate transformation, the rest observer at the spatial origin in the coordinate (\ref{coor}) changes to a moving observer.
% in the new coordinate system.
The new coordinate corresponds to the galactic coordinate if $\Lambda =0$, and it is a convenient coordinate for various astrophysical models.
%%We include the lensing objects to the de Sitter spacetime in the perturbative expression (\ref{sdS}).
%%As the same way in Eq.(\ref{Eeq}), we obtain a perturbative solution of Einstein equation
%In this coordinate system, it is the rest observer at the spatial origin ($\vec{r}_o=\vec{0}$) that captures the cosmic expansion correctly. We cannot follow the causal structure of any other observers in this coordinate. This reflects that the cosmic expansion is observer-dependent.
%We want to describe the cosmic expansion seen by a slowly moving observer so that we change the spatial coordinate of the observer from the rest one $\vec{r}_o=\vec{0}$ to the moving one (\ref{ro}), $\vr_o = \vr_{o,{\rm ini}} + \vv_o t$.
%
We can change to the new coordinate by the following transformation:
\begin{align}
%\tilde{x}^\mu = \Lambda^\mu_\nu (x^\nu - \delta^\nu_i r_{B,{\rm ini}}^i) \\
\tilde{t}& = \gamma_B[t - \vv_B \cdot (\vr - \vr_{B,{\rm ini}})]\ , \nonumber \\
\vec{\tilde{r}}&= \vr - \vr_o + \frac{(\gamma_B-1)}{v_B^2}\vv_B\cdot (\vr - \vr_B) \vv_B\ ,\nonumber \\
v_B&:= |\vv_B|, \quad \gamma_B:= 1/\sqrt{1-v_B^2}\ .
\end{align}
This consists of a spatial translation and a boost in terms of the coordinate system.
The rest observer at the spatial origin in the coordinate (\ref{coor}) transforms to
\begin{align}
\vec{\tilde{r}}_o& = -\tilde{\vec{r}}_{o,{\rm ini}} - \vv_B\tilde{t}, \nonumber \\
\tilde{\vec{r}}_{o,{\rm ini}}&:= \vr_{B,{\rm ini}} +\left(\gamma_B -\frac{\gamma_B-1}{v_B^2}\right)
(\vv_B\cdot \vr_{B,{\rm ini}})\vv_B\ .
\end{align}
We obtain the following metric in the new coordinate system up to the first order of perturbation:
\begin{align}
\label{isot}
ds^2 &= -\left(1-\sum_J\frac{2GM_J}{|\vec{\tilde{r}}-\vec{\tilde{r}}_J|} -\frac{1}{3}\Lambda
|\vec{\tilde{r}}-\vec{\tilde{r}}_o|^2\right) d\tilde{t}^2 \nonumber \\
&\quad + \left(1+\sum_J\frac{2GM_J}{|\vec{\tilde{r}}-\vec{\tilde{r}}_J|} -\frac{1}{6}\Lambda
|\vec{\tilde{r}}-\vec{\tilde{r}}_o|^2\right)
d\vec{\tilde{r}}^2 \ . %(d\tilde{r}^2+\tilde{r}^2d\Omega_2) \ .
\end{align}
Even in the new coordinate system, the metric remains spatially conformal flat up to the first order of perturbation, and the effect of the cosmological constant vanishes at the moving observer.
%%In the new coordinate (\ref{coor}), the moving observer who captures the cosmic expansion correctly has the apparent velocity $\vv_o$ relative to a rest observer who cannot capture the cosmic expansion. Also, the light entering the world point of the moving observer has the apparent velocity (\ref{ve}) relative to the rest observer. To define observables seen by the moving observer, we have to boost the system to the local rest frame of the moving observer as in Eq.(\ref{boost}).

Then, we calculate and compare the angle differences observed at $\vec{\tilde{r}}_o$ and $\vr_o$ in the both coordinates. In the coordinate (\ref{isot}), the observer moves slowly with his/her velocity $-\vv_B$ so that we need the Lorentz boost to the local rest frame to get observables. For the small angle differences, we obtain
\begin{align}
\label{dec1}
&\widetilde{\rm DEC}_{ij}\simeq \frac{\tilde{k}^{\hat{x}\rq{}}_{s_i}}{\tilde{k}^{\hat{z}\rq{}}_{s_i}}
-\frac{\tilde{k}^{\hat{x}\rq{}}_{s_j}}{\tilde{k}^{\hat{z}\rq{}}_{s_j}}, \\
&\tilde{k}^{\hat{\alpha}\rq{}}_{s_i} = \Lambda^{\hat{\alpha}}_{B\hat{\beta}}
{\tilde{e}^{\hat{\beta}}_{\ \mu}\frac{\partial \tilde{x}^\mu}{\partial x^\rho}k^{\rho}_{s_i}}, \nonumber \\
% &\frac{\tilde{k}^{\hat{x}\rq{}}_{s_i}}{\tilde{k}^{\hat{z}\rq{}}_{s_i}}=
% \frac{\tilde{e}^{\hat{x}\rq{}}_{\ \mu} \tilde{k}^{\mu}_{s_i}}
% {\tilde{e}^{\hat{z}\rq{}}_{\ \nu} \tilde{k}^{\nu}_{s_i}}
% = \frac{\tilde{e}^{\hat{x}}_{\ \mu} \frac{\partial \tilde{x}^\mu}{\partial x^\rho}k^{\rho}_{s_i}}
% {\tilde{e}^{\hat{z}}_{\ \nu} \frac{\partial \tilde{x}^\nu}{\partial x^\sigma}k^{\sigma}_{s_i}}, \nonumber \\
&\Lambda^{\hat{\alpha}}_{B\hat{\beta}} = \left(
\begin{array}{cc}
\hat{\gamma}_B & \hat{\gamma}_B v_B^{\hat{j}} \\
\hat{\gamma}_B v^{\hat{i}}_B & \delta_{\hat{i}\hat{j}}
+ (\hat{\gamma}_B-1)\frac{v^{\hat{i}}_Bv^{\hat{j}}_B}{\hat{v}_B^2}
\end{array}
\right) \ , \nonumber \\
&v_B^{\hat{i}}= \sqrt{|\tilde{g}^{00}|\tilde{g}_{ij}}v_B^i,\quad
\hat{\gamma}_B:= 1/\sqrt{1-v_B^{\hat{i}}v_B^{\hat{i}}}, \nonumber \\
&{\tilde{e}^{\hat{0}}}_{\ \mu} = (1/\sqrt{|\tilde{g}^{00}|},\ 0,\ 0,\ 0), \nonumber \\
&{\tilde{e}^{\hat{1}}}_{\ \mu} = (0,\ \sqrt{\tilde{g}_{11}},\ 0,\ 0), \nonumber \\
&{\tilde{e}^{\hat{2}}}_{\ \mu} = (0,\ 0,\ \sqrt{\tilde{g}_{22}},\ 0), \nonumber \\
&{\tilde{e}^{\hat{3}}}_{\ \mu} = (0,\ 0,\ 0,\ \sqrt{\tilde{g}_{33}}), \nonumber
% &\tilde{e}^{\hat{\alpha}}_{\ \mu} = {\rm diag}(\sqrt{|\tilde{g}_{00}|},\sqrt{\tilde{g}_{11}}
% ,\sqrt{\tilde{g}_{22}},\sqrt{\tilde{g}_{33}}) \ .
\end{align}
where all the quantities with the tilde are defined using the new coordinate (\ref{isot}), and $k^\rho_{s_i}$ is four-momentum of the light in the original coordinate (\ref{coor}).
%\begin{align}
% \tilde{DEC}_{si}&:= -\frac{\tilde{k}^{\hat{x}}_{s_i}}{\tilde{k}^{\hat{z}}_{s_i}}
% = -\frac{\tilde{e}^{\hat{x}}_{\ \mu} \tilde{k}^{\mu}_{s_i}}
% {\tilde{e}^{\hat{z}}_{\ \nu} \tilde{k}^{\nu}_{s_i}} \\
% &= - \frac{\tilde{e}^{\hat{x}}_{\ \mu} \frac{\partial \tilde{x}^\mu}{\partial x^\rho}k^{\rho}_{s_i}}
% {\tilde{e}^{\hat{z}}_{\ \nu} \frac{\partial \tilde{x}^\nu}{\partial x^\sigma}k^{\sigma}_{s_i}}
%\end{align}
%
While, the small angle differences calculated in the original coordinate (\ref{coor}) is expressed as
\begin{align}
\label{dec2}
&{\rm DEC}_{ij}\simeq \frac{k^{\hat{x}}_{s_i}}{k^{\hat{z}}_{s_i}}
-\frac{k^{\hat{x}}_{s_j}}{k^{\hat{z}}_{s_j}}, \\
&k^{\hat{\alpha}}_{s_i}= e^{\hat{\alpha}}_{\ \rho}k^{\rho}_{s_i} \ . \nonumber
% \frac{\Lambda^{\hat{x}}_{\hat{\alpha}}{e^{\hat{\alpha}}}_{\rho} k^{\rho}_{s_i}}
% {\Lambda^{\hat{z}}_{\hat{\beta}}{e^{\hat{\beta}}}_{\sigma} k^{\sigma}_{s_i}}.
\end{align}
Comparing Eq.(\ref{dec1}) and (\ref{dec2}), we can easily find that they correspond each other up to the first order of perturbation since we approximately hold the following relation
\begin{equation}
\Lambda^{\hat{\alpha}}_{B\hat{\beta}}\tilde{e}^{\hat{\beta}}_{\ \mu} \frac{\partial \tilde{x}^\mu}{\partial x^\rho}
\simeq{e^{\hat{\alpha}}}_{\rho} \ .
\end{equation}
We obtain the same result for $\widetilde{\rm RA}_{ij}$, and we can conclude that
the angle differences defined in Eq.(\ref{dangle}) are the coordinate invariant observables at least up to the first order of perturbation.
We comment that by definition, the angle differences calculated in different coordinates will correspond each other even in a non-perturbative regime.

%%%%%%%%%%%%%%%%%%%%%%%%%%%%%%%%%%%%%%%%%%%%%%%%%%%%%%%%%%%%%%
\bibliographystyle{apsrmp}
%\bibliography{rmp-sample}

\end{document}